\documentclass{sf2a-conf}
\usepackage{graphicx}
\usepackage{natbib}
\begin{document}

\TitreGlobal{SF2A 2008}

\title{How Polarization and Scattering can reveal Geometries, Dynamics, and Feeding of Active Galactic Nuclei}
\author{Goosmann, R.~W.}\address{Observatoire Astronomique de Strasbourg, 11 rue de l'Universit\'e, F-67000 Strasbourg, France}
\author{Gaskell, C.~M.}\address{Department of Astronomy, University of Texas, Austin, TX 78712-0259, USA}
\author{Shoji, M.$^2$}
\runningtitle{Polarization and Scattering in AGN}
\setcounter{page}{237}

\index{Goosmann, R.~W.}
\index{Gaskell, C.~M.}
\index{Shoji, M.}

\maketitle
\begin{abstract}
\Citet{gaskell2007} introduced polarization reverberation mapping as a new technique to explore the structure of active galactic nuclei. We present modeling results for the time-dependent polarization signal expected from scattering inside a centrally illuminated spheroid. Such a model setup describes a larger corona surrounding the compact source of an active nucleus. Time-delays between the polarized and the total flux are computed and related to the geometry of the cloud and the viewing angle. When including the in-flow dynamics of the cloud, it is possible to constrain its optical depth and velocity, which enables estimations of the mass inflow rate. 
\end{abstract}
%

\section{Introduction}

Multiple reprocessing and scattering in the central engine of active galactic nuclei (AGN) are a major challenge to the spectral analysis. To disentangle the various media present in the inner parts of an AGN, the radiative transfer has to be modeled coherently. Scattering induces polarization that can be exploited to constrain the emission and reprocessing geometry. The Monte-Carlo radiative transfer code {\sc stokes} \citep{goosmann2007} is designed to model polarization in a wide range of astronomical applications. In its most recent version, time-dependent effects are included, which allows to interpret results from polarization reverberation mapping \citep{gaskell2007}. In this proceedings note we present {\sc stokes} reverberation modeling for the polarized echo of a centrally illuminated scattering cloud. 

\section{Constraining the shape of a spheroidal scattering cloud}

We model the time-resolved, polarized flux of a centrally illuminated, constant density, and Thomson scattering spheroid. The vertical half-axis of the spheroid is denoted by $a$, the horizontal half-axis by $b$. We define $b = 10$~light-days and evaluate the aspect ratios $a/b = 1/10$,~$3/10$,~$5/10$, and~$7/10$ at viewing angles of $i = 10^\circ$,~$30^\circ$, and~$60^\circ$ (measured from the spheroid's symmetry axis). The Thomson optical depth $\tau$ along $b$ is set to unity.

\begin{figure}[t]
   \centering
   \includegraphics[width=\textwidth]{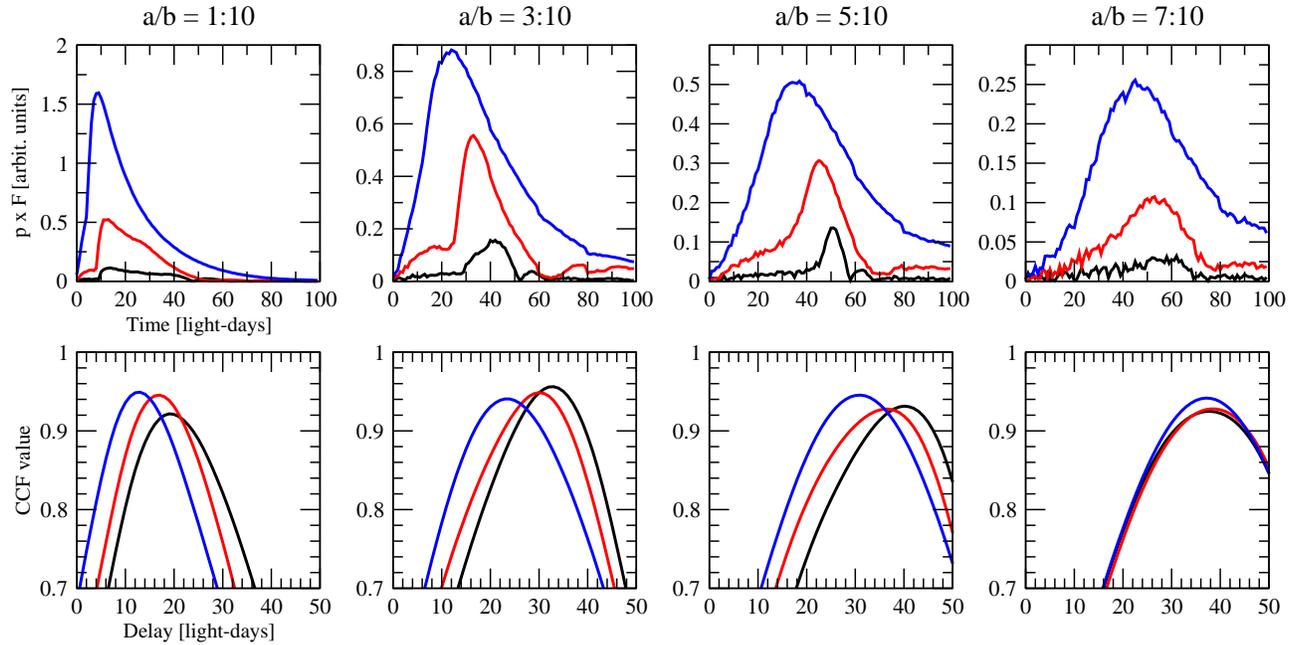}
   \caption{Top: lightcurves for the polarized flux of a light flash emitted at the center of a spheroidal scattering region with aspect ratio $a/b$. The viewing angles $i = 10^\circ$ (black), $30^\circ$ (red), and $60^\circ$ (blue) are evaluated. Bottom: cross-correlation functions between the total flux and the polarized flux echo when the intensity of the central source varies.}
   \label{fig1}
\end{figure}

At first, we assume a short flash of unpolarized light to be emitted at the center of the spheroid. We trace for each viewing angle the time evolution of the escaping radiation. The resulting lightcurves of the polarized flux, i.e. the percentage of polarization $p$ multiplied by the total flux $F$, are shown in the top row of Fig.~\ref{fig1}. With increasing $a/b$ the maxima of the curves shift toward higher values and their shape becomes more symmetric and shallow. For a given $a/b$, the normalization of the lightcurves increases with $i$ while the position of the maximum moves toward shorter delays. Except for the flattest spheroid, the light-crossing distances indicated by the maxima of the lightcurves exceed the horizontal extension (20 light-days) of the spheroid. This clearly indicates multiple scattering effects.

Next, we randomly vary the intensity of the central source and conduct a cross-correlation analysis between $p \times F$ and $F$. The results are shown in the bottom row of Fig.~\ref{fig1}. The correlation between the total and the polarized flux is always strong. The resulting delays depend on $a/b$ and $i$ and the dependencies correspond to those of the single light flashes studied before. Note that a degeneracy appears for the dependence on $i$ when the shape of the scattering cloud approaches a sphere (see the case of $a/b = 7/10$). This, of course, is expected from symmetry considerations. 

\section{Determining the optical depth and velocity of in-falling matter}

\begin{figure}[h]
  \begin{minipage}{0.60\textwidth} 
    \caption{Calculated shifts of an emission line by electron or Rayleigh scattering \citep[see][for details of the computation]{gaskell2008}. The top black solid curve shows a Lorentzian line profile before scattering. The other solid curves show the blue-shifting caused by an external spherical shell of scatterers with an inflow velocity of -1000 km/s. In order of increasing blue-shifting and decreasing peak flux, the curves show the effects of $\tau$ = 0.5 (red), 1.0 (green), 2.0 (blue), and 10 (purple). The dots show the blue-shifting caused by a cylindrical distribution of scatters with $\tau = 20$ inflowing at -1000 km/s.}
    \label{fig2}
  \end{minipage}
  \hfill
  \begin{minipage}{0.35\textwidth}
    \vskip -0.4 truecm
    \centering
    \includegraphics[width=\textwidth]{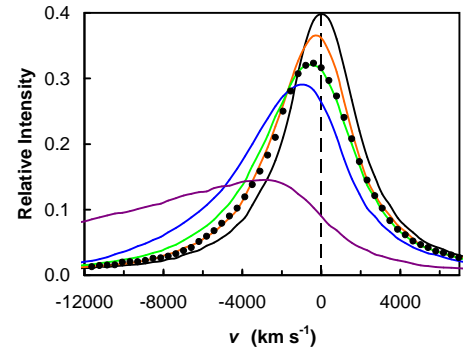}
  \end{minipage}
\end{figure}

The correlations shown in Fig.~\ref{fig1} could be used to constrain the geometry of the scattering cloud if the inclination of the system is known. But the time-delays of the polarization echo also depend on $\tau$. To put constraints on $\tau$ the dynamics of the scattering medium can be exploited, as was recently shown in \citet{gaskell2008} for scattering inflows in AGN. In Fig.~\ref{fig2} we show the effects of multiple scattering inside an in-falling medium. The resulting blue-shifting of emission lines is observed in AGN, and our modeling with {\sc stokes} has shown that the shifting and the resulting line profile do not strongly depend on the geometry of the inflow but mainly on its velocity and optical depth. These quantities are closely related to the inflow rate of the material and thus to the possible growth rate of the supermassive black hole.

\end{document}